\documentclass[a4paper,12pt]{article}
\usepackage{amssymb}
\usepackage{amsfonts}
\usepackage{amsmath}
\usepackage{epsfig}
\usepackage{color}
\usepackage{accents}
\usepackage{hyperref}

\newcommand{\be}{\begin{equation}}
\newcommand{\ee}{\end{equation}}
\newcommand{\bd}{\begin{displaymath}}
\newcommand{\ed}{\end{displaymath}}
\newcommand{\bea}{\begin{eqnarray}}
\newcommand{\eea}{\end{eqnarray}}

\begin{document}
\begin{titlepage}

\baselineskip 24pt
\newcommand{\sheptitle}
{\Large Estimating strong decays of $X(3915)$ and $X(4350)$}

\newcommand{\shepauthor}
{Zi-Jia Zhao$^{1*}$, Dong-Mei Pan$^1$}

\newcommand{\shepaddress}
{$^1$Department of Modern Physics,
University of Science and Technology of China\\Hefei, Anhui 230026, China}

\newcommand{\shepabstract}
{Strong decays of $X(3915)$ and $X(4350)$ have been studied by assuming them as P-wave charmonium  states. We estimate the  $\Gamma(X(3915)\rightarrow J/\psi \omega)$ and $\Gamma(X(4350)\rightarrow J/\psi \phi)$ via $D \bar{D}^{(*)}$ open-charm intermediate states. Our calculation supports that the assignment of $X(4350)$ to the charmonium state while the assignment of $X(3915)$ is disfavored.}


\begin{flushright}
\end{flushright}

\begin{center}
{\large{\bf \sheptitle}}
\bigskip \\ \mbox{} \\{\large \shepauthor}\\ {\it \shepaddress} \\ \vspace{.5in}
{\bf Abstract} \bigskip \end{center} \setcounter{page}{0}
\shepabstract
\begin{flushleft}
\end{flushleft}
\vfill \noindent

$^*$ E-mail: ~sszdzl1@mail.ustc.edu.cn
\end{titlepage}

\section{Introduction}

Two new charmoniumlike  resonances named $X(3915)$ and $X(4350)$ were  observed recently by Belle Collaboration \cite{Belle 1, Belle 2} in the processes $\gamma\gamma\rightarrow J/\psi \omega$ and $\gamma\gamma\rightarrow J/\psi \phi$, respectively. For $X(3915)$, $M=3915\pm 3(\rm stat)\pm 2{\rm syst}$ MeV and $\Gamma=17\pm10(\rm stat)\pm3(\rm syst)$MeV; for $X(4350)$, $M=4350.6_{-5.1}^{+4.6}(\rm stat)\pm0.7(\rm syst)$ MeV and $\Gamma=13_{-9}^{+18}(\rm stat)\pm4(\rm syst)$ MeV.  The structure of these states has been studied extensively in the literature, the $P$-wave charmonium states \cite{Xiang Liu 1,0912.5061}, the $D^*_s$ and $D^*_{s0}$ molecule state \cite{1006.1276}, The $c\bar{c} s\bar{s}$ teraquark state \cite{Stancu2010}, and the scalar $c\bar{c}$ and $D^*_s \bar{D}^*_s$ mixing state \cite{ZGWang2010} and so on.

The purpose of the present paper is to further study strong decays $X(3915)\rightarrow J/\psi \omega$ and $X(4350)\rightarrow J/\psi \phi$, in order to increase our understanding in the structure of these new resonances. We will treat them as P-wave charmonium states, following Ref. \cite{Xiang Liu 1}, and the $J^{PC}$ quantum numbers of theirs are $0^{++}$ for $X(3915)$ and $2^{++}$ for $X(4350)$. Thus their open-charm decay amplitudes, such as  $X(3915)\rightarrow D \bar{D}$ and $X(4350)\rightarrow D_s^{(*)}\bar{D}_s^{(*)}$ transitions,  can be easily obtained using the 3p0 model \cite{book}. It is obvious that these open-charm intermediate states may rescatter into $J/\psi\omega$ or $J/\psi \phi$ final states, and this rescattering effects can be captured using the method \cite{cheng} used in \cite{0904.0316,0610.278}.

The paper is organized as follows: In Sec. 2 and Sec. 3, we present the formalism used in our study, and give explicit calculations for $X(3915)$ and $X(4350)$. We summarize our results in Sec. 4. The open-charm decay amplitudes of $X(3915)$ and $X(4350)$ are fixed in the Appendix by 3p0 model. This will help us to determine the coupling constants of $X(3915)\rightarrow D\bar{D}$ and $X(4350)\rightarrow D_s^{(*)}\bar{D}_s^{(*)}$, which will be used in the calculations of Sec. 2 and Sec. 3.

\section{Calculation for X(3915)}
The process $X(3915)\rightarrow J/\psi \omega$ is OZI rule suppressed,  so the final state interaction (FSI) effects
may play the central role. We will study if the hidden charm decay
$X(3915)\rightarrow J/\psi \omega$ mainly arises from the FSI effect
of $X(3915)\rightarrow D^0 \bar{D}^0$ and $X(3915)\rightarrow D^+
D^-$ rather than $X(3915)\rightarrow D^{*0} \bar{D}^{*0}$ or
$X(3915)\rightarrow D^{*+} \bar{D}^{*-}$ because $D^{*}$'s are too
heavy.

The strong interactions between X(3915) and $D$'s can be described
by the following phenomenological Lagrangian:
\begin{eqnarray}
{\cal L}_{0^+DD}=g_{0^+D D} (X D^0 \bar{D}^0+ X D^+ D^-)
\end{eqnarray}

The strong coupling constants $g$ can be calculated by some physical
models, for example, the 3p0 model, which will be
shown in the Appendix, the numeral result is $|g_{0^+D D}| = 2760
MeV$.

The Feynman diagrams for $X(3915) \rightarrow J/\psi \omega$
through $D^+$ and $D^-$ is depicted in Figure 1.
\begin{figure}[htb]
\centering \scalebox{0.7}{\includegraphics{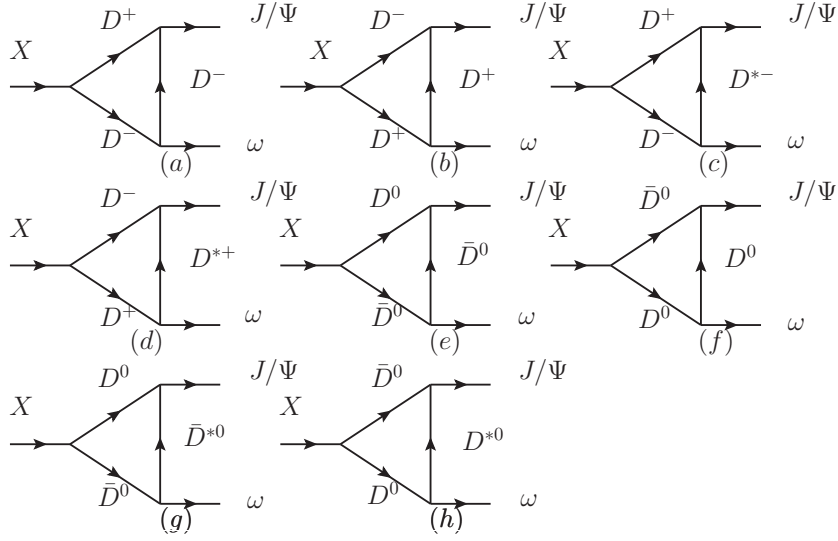}} \caption{The
diagrams for $X(3915)\rightarrow J/\psi+\omega$ via $D^+ D^-$ and
$D^0 \bar{D^0}$}
\end{figure}

 Based on the heavy quark symmetry
and chiral symmetry, the effective Lagrangian was constructed
as\cite{prc 63.034901}\cite{prd 46.1148}\cite{pr 281.145}:
\begin{eqnarray}
{\cal L}_{J/\psi D D} = \textit{i} \textit{g}_{J/\psi D D}
\psi_\mu(\partial^\mu D D^\dagger-D \partial^\mu D^\dag)
\end{eqnarray}
\begin{eqnarray}
{\cal L}_{J/\psi D D^*}=-\textit{g}_{J/\psi D D^*}
\varepsilon^{\mu\nu\alpha\beta} \partial_\mu\psi_\nu(\partial_\alpha
D^*_\beta D^\dag + D\partial_\alpha D^{*\dag}_\beta)
\end{eqnarray}
\begin{eqnarray}
{\cal L}_{J/\psi D^* D^*}=&&-\textit{i}\textit{g}_{J/\psi D^* D^*}\{\psi^\mu(\partial_{\mu} D^{*\nu}D^{*\dag}_\nu-D^{*\nu}\partial_\mu D_\nu^{*\dag})+(\partial_\mu\psi_\nu D^{*\nu}-\psi_\nu\partial_\mu D^{*\nu})D^{*\mu\dag}\nonumber\\
&&+D^{*\mu}(\psi^\nu\partial_\mu D^{*\dag}_\nu-\partial_\mu\psi_\nu D^{*\nu\dag})\}
\end{eqnarray}
\begin{eqnarray}
{\cal
L}_{DDV}=-\textit{i}\textit{g}_{DDV}D^\dag_\textit{i}\overleftrightarrow\partial_\mu
D^\textit{j}(V^\mu)^\textit{i}_\textit{j}
\end{eqnarray}
\begin{eqnarray}
{\cal
L}_{D^*DV}=-\textrm{2}\textit{f}_{D^*DV}\varepsilon_{\mu\nu\alpha\beta}(\partial
V^\nu)^\textit{i}_\textit{j}(D^\dag_\textit{i}\overleftrightarrow\partial^\alpha
D^{*\beta\textit{j}}-D^{*\beta\dag}_\textit{i}\overleftrightarrow\partial^\alpha
D^\textit{j})
\end{eqnarray}
\begin{eqnarray}
{\cal
L}_{D^*D^*V}=\textit{i}\textit{g}_{D^*D^*V}D^{*\nu\dag}_\textit{i}\overleftrightarrow\partial_\mu
D^{*\textit{j}}_\nu(V^\mu)^\textit{i}_\textit{j}+\textrm{4}\textit{i}\textit{f}_{D*D*V}D^{*\dag}_{\textit{i}\mu}(\partial^\mu
V^\nu-\partial^\nu V^\mu)^\textit{i}_\textit{j}D^{*\textit{j}}_\nu
\end{eqnarray}

$V$ denote the nonet vector meson matrices
\begin{displaymath}
\mathbf{V} = \left( \begin{array}{ccc}
\frac{\rho^0}{\sqrt{2}}+\frac{\omega}{\sqrt{2}} & \rho^+ & K^{*+} \\
\rho^- & -\frac{\rho^0}{\sqrt{2}}+\frac{\omega}{\sqrt{2}} & K^{*0} \\
K^{*-} & \overline K^{*0} & \phi
\end{array} \right)
\end{displaymath}

 The value of the coupling constants are \cite{pr 281.145} $g_{DDV}=g_{D^*D^*V}=\beta g_V/\sqrt{2}$ , $g_V=m_\rho/f_\pi$, $f_{\pi}=132 MeV$, $f_{D^*DV}=f_{D^*D^*V}/m_{D^*}=\lambda g_V/\sqrt{2}$, we take $\beta=0.9$, and $\lambda=0.56 GeV^{-1}$ from \cite{prd 68.114001}. $g_{J/\psi DD}=\sqrt{20\pi}$, $g_{J/\psi D*D*}=g_{J/\psi DD}=m_D g_{J/\psi DD^*}$ which was determined in \cite{pr 281.145} from the chiral and heavy quark limit.

 By the Cutkosky cutting rule, the absorptive part of Fig.1 is written
 as
 \begin{eqnarray}
 A_{1-a}=&&\frac{1}{2}\int \frac{d^3 p_1}{(2\pi)^3 2E_1}\frac{d^3p_2}{(2\pi)^3 2E_2}\delta^4(m_X-p_1-p_2)(2\pi)^4 ig_{0^+DD}(-i)g_{J/\psi DD}\,p_1\cdot\varepsilon_{\psi}\nonumber\\
 &&(-\frac{i}{\sqrt{2}})g_{DDV}\, \varepsilon_\omega\cdot p_{2}\frac{i}{q^2-m_D^2}{\cal
 F}^2(M_{D^-},q)
 \end{eqnarray}
 \begin{eqnarray}
 A_{1-c}=&&\frac{1}{2} \int \frac{d^3 p_1}{(2\pi)^3 2E_1}\frac{d^3p_2}{(2\pi)^3 2E_2}\delta^4(m_X-p_1-p_2)(2\pi)^4 ig_{0^+DD}(-i) g_{J/\psi DD^*}p_\psi^\mu \varepsilon_\psi^\nu q_\alpha \epsilon_{\mu\nu\alpha\beta}\nonumber\\
 &&(-\sqrt{2}i)f_{D^*DV} \epsilon_{\rho\sigma\kappa\eta}(-i p_\omega^\rho)\varepsilon_\omega^\sigma(-i)(-g^{\beta\eta}+\frac{q^\beta q^\eta}{m_{D^*}^2})(q^\kappa+p^\kappa_2)\frac{i}{q^2-m_{D^*}^2}{\cal F}^2(M_{D^{*-},q})
 \end{eqnarray}
 \begin{eqnarray}
  A_{1-e}=&&\frac{1}{2}\int \frac{d^3 p_1}{(2\pi)^3 2E_1}\frac{d^3p_2}{(2\pi)^3 2E_2}\delta^4(m_X-p_1-p_2)(2\pi)^4 if_{0^+DD}(-i)g_{J/\psi DD}\,p_1\cdot\varepsilon_{\psi}\nonumber\\
 &&(-\frac{i}{\sqrt{2}})g_{DDV}\, \varepsilon_\omega\cdot p_{2}\frac{i}{q^2-m_D^2}{\cal F}^2(M_{\bar{D}^0},q)
 \end{eqnarray}
 \begin{eqnarray}
 A_{1-g}=&&\frac{1}{2} \int \frac{d^3 p_1}{(2\pi)^3 2E_1}\frac{d^3p_2}{(2\pi)^3 2E_2}\delta^4(m_X-p_1-p_2)(2\pi)^4 if_{0^+DD}(-i) g_{J/\psi DD^*}p_\psi^\mu \varepsilon_\psi^\nu q_\alpha \epsilon_{\mu\nu\alpha\beta}\nonumber\\
 &&(-\sqrt{2}i)f_{D^*DV} \epsilon_{\rho\sigma\kappa\eta}(-i p_\omega^\rho)\varepsilon_\omega^\sigma(-i)(-g^{\beta\eta}+\frac{q^\beta q^\eta}{m_{D^*}^2})(q^\kappa+p^\kappa_2)\frac{i}{q^2-m_{D^*}^2}{\cal F}^2(M_{\bar{D}^0},q)
 \end{eqnarray}
 where $ {\cal F}^2(m_i,q)= (\frac{\Lambda^2-m_i^2}{\Lambda^2-q^2})^{2}$, is a form factor which compensate
the off-shell effects of mesons at the vertices and are normalized
at $q^2=m_i^2$. $q=p_\psi-p_1=p_2-p_\omega$, $\Lambda(m_i)=m_i +
\alpha \Lambda_{QCD} $, $m_i$ denote the mass of the exchanged
particle and we choose $\Lambda_{QCD}=220 MeV$ here. $\alpha$ should
not be far from 1. So the total decay amplitude of
$X(3915)\rightarrow J/\psi \omega$ is:
 \begin{eqnarray}
 {\cal M}(X(3915)\rightarrow D^+ D^- \rightarrow J/\psi \omega)=2(A_{1-a}+A_{1-c}+A_{1-e}+A_{1-g})
 \end{eqnarray}

The factor 2 comes form the fact that the amplitudes of Figure 1-b, 1-d, 1-f, 1-h are the same with Figure 1-a, 1-c, 1-e, 1-g.
The relation between decay width of X(3915) and $\alpha$ is shown in Figure.2.
 \begin{figure}[htb]
\centering \scalebox{0.8}{\includegraphics{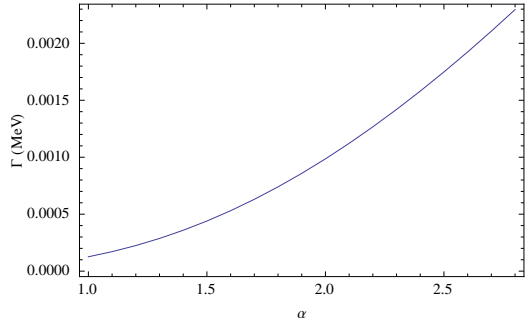}}
\caption{The decay rate for $X(3915)\rightarrow J/\psi+\omega$ via $D^+ D^-$ and $D^0 \bar{D^0}$}
\end{figure}

\section{Calculation for X(4350)}

 Similarly, the hidden charm decay $X(4350)\rightarrow J/\psi \phi$, which is OZI rule suppressed, may occur through $D_s^{+}$, $D_s^{-}$ and $D_s^{+*}$, $D_s^{+*}$ rescattering.\\
 The strong interaction between X(4350)and $D_s$, $D^*_s$ can be described in a phenomenological way:
\begin{eqnarray}
{\cal L}_{2^+DD}=g_{2^+D D} X^{\mu\nu} \partial_\mu D_s^+
\partial_\nu D_s^-
\end{eqnarray}
\begin{eqnarray}
{\cal L}_{2^+D^*D}=ig_{2^+D^*D} \partial^{\mu}X^{\nu\alpha}\varepsilon_{\mu\nu\beta\rho}(\partial^\beta D_s^{*+\rho} \partial_\alpha D_{s}^- +\partial^\beta D_s^{*-\rho} \partial_\alpha D_{s}^{+})
\end{eqnarray}
\begin{eqnarray}
{\cal L}_{2^+D^*D^*}=g_{2^+D^*D^*} X^{\mu\nu} D_{s\mu}^{*+}D_{s\nu}^{*-}
\end{eqnarray}
 The absolute value of $g_{2^+DD}$, $g_{2^+D^*D}$, $g_{2^+D^*D^*}$ can be decided by 3p0 model which will be shown in Appendix, the numeral result is $|g_{2^+DD}|=0.002 MeV^{-1}$, $|g_{2^+D^*D}|=3.30\times 10^{-7} MeV^{-2}$, $|g_{2^+D^*D^*}|=700 MeV$. The Feynman diagrams for $X(4350)\rightarrow J/\psi+\phi$ are shown in Figure.3
\begin{figure}[htb]
\centering \scalebox{0.6}{\includegraphics{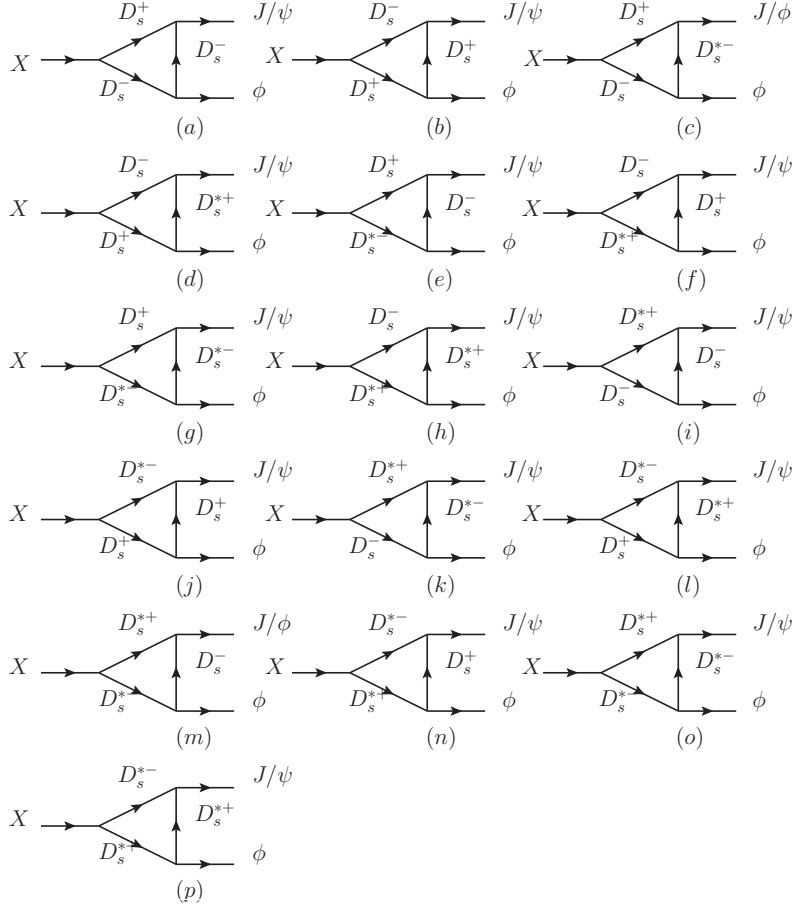}}
\caption{The diagrams for $X(4350)\rightarrow J/\psi+\phi$ via $D_s$ and $D_s^*$ assuming $X(4350)$ is $\chi''_{c2}$}
\end{figure}

 The absorptive part of Fig.2 is written as

\begin{eqnarray}
 A_{2-a}=&&\frac{1}{2} \int \frac{d^3 p_1}{(2\pi)^3 2E_1}\frac{d^3p_2}{(2\pi)^3 2E_2}\delta^4(m_X-p_1-p_2)(2\pi)^4 (-i)g_{2+DD}p_{1\mu} p_{2\nu} \varepsilon_X^{\mu\nu}\nonumber\\
 &&(-ig_{J/\psi DD})\varepsilon_\alpha^*(\psi)(p_1^\alpha-q^\alpha)(-i)g_{DDV}(q_\beta+p_{2\beta})\xi_\phi^*\frac{i}{q^2-m_{D_s}^2} {\cal F}^2(M_{D_s^{-}},q)
\end{eqnarray}

\begin{eqnarray}
 A_{2-c}=&&\frac{1}{2}\int \frac{d^3 p_1}{(2\pi)^3 2E_1}\frac{d^3p_2}{(2\pi)^3 2E_2}\delta^4(m_X-p_1-p_2)(2\pi)^4 i g_{2^+DD} \xi_{X}^{\kappa\varphi}p_{1\kappa}p_{2\varphi} \nonumber\\
 &&g_{J/\psi DD^*}\epsilon_{\mu\nu\alpha\beta}p_\psi^\mu \xi_\psi^{*\nu} p_1^{\alpha} (-g^{\beta\eta}+\frac{q^\beta q^\eta}{m_{D_s^*}^2})f_{D^*DV}\epsilon_{\rho\sigma\tau\eta} p_\phi^\rho \varepsilon_\phi^{*\sigma}2(p_2^\tau+q^\tau) \nonumber\\
 &&\frac{i}{q^2-m^2_{D_s^*}}{\cal F}^2(M_{D_s^{*-}},q)
\end{eqnarray}

\begin{eqnarray}
A_{2-e}=&&\frac{1}{2}\int \frac{d^3 p_1}{(2\pi)^3 2E_1}\frac{d^3p_2}{(2\pi)^3 2E_2}\delta^4(m_X-p_1-p_2)(2\pi)^4 (-i)g_{2^+D^*D} \xi_X^{\kappa\eta} p_X^\rho \epsilon_{\rho\kappa\tau\sigma} p_2^\tau p_{1\eta} \nonumber\\
&&ig_{J/\psi DD}\xi_\psi^*\cdot (p_1-q)(-2i)
f_{D^*DV}\xi_{\mu\nu\alpha\beta}\xi_\phi^{\nu*}p_{\phi}^{\mu}(p_2^\alpha+q^\alpha)(-g^{\sigma\beta}+\frac{p_2^\sigma
p_2^\beta}{m_{D_s^*}^2}) \nonumber\\
&&\frac{i}{q^2-m_{D_s}^2}{\cal F}^2(M_{D_s^{-}},q)
\end{eqnarray}

\begin{eqnarray}
A_{2-g}=&&\frac{1}{2}\int \frac{d^3 p_1}{(2\pi)^3 2E_1}\frac{d^3p_2}{(2\pi)^3 2E_2}\delta^4(m_X-p_1-p_2)(2\pi)^4 (-i) g_{2^+D^*D}p_X^\eta \varepsilon_X^{\rho\varphi}\epsilon_{\eta\rho o\sigma}p_2^o p_{1\varphi}\nonumber\\
&&ig_{J/\psi DD^*}\epsilon^{\tau\kappa\gamma\beta}p_{\psi\tau}\varepsilon^{*}_{\psi\kappa}p_{1\gamma}[-g_{D^*D^*V}(ip_{2\mu}+iq_{\mu}) \varepsilon_\phi^{*\mu} g^{\alpha\nu}-4f_{D^*D^*V}(-ip_\phi^\alpha\varepsilon_\phi^{*\nu}+ip_\phi^\nu\varepsilon_{\phi}^{*\alpha})]\nonumber\\
&&(-g_{\ \nu}^\sigma +\frac{p_{2}^\sigma p_{2\nu}}{m_{D_s^*}})(-g_{\beta\alpha}+\frac{q_{\beta}q_{\alpha}}{m_{D_s^*}})\frac{i}{q^2-m_{D_s^*}^2}{\cal F}^2(M_{D_s^{*-}},q)
\end{eqnarray}

\begin{eqnarray}
A_{2-j}=&&\frac{1}{2}\int \frac{d^3 p_1}{(2\pi)^3 2E_1}\frac{d^3p_2}{(2\pi)^3 2E_2}\delta^4(m_X-p_1-p_2)(2\pi)^4 (-i)g_{2^+D^*D} p_X^\eta \varepsilon_X^{\rho o} \epsilon_{\eta\rho\kappa\sigma} p_1^\kappa p_{2 o}\nonumber\\
&&(-i)g_{J/\psi DD^*} \epsilon^{\mu\nu\alpha\beta}(-i)p_{\psi\mu} \varepsilon_{\psi\nu}^*(ip_{1\alpha})2i g_{DDV} p_{2\tau} \varepsilon_\phi^{*\tau}(-g^\sigma_{\ \beta}+\frac{p^\sigma_1 p_{1\beta}}{m_{D_s^*2}})\nonumber\\
&&\frac{i}{q^2-m_{D_s^*}^2}{\cal F}^2(M_{D_s^{+}},q)
\end{eqnarray}

\begin{eqnarray}
A_{2-l}=&&\frac{1}{2}\int \frac{d^3 p_1}{(2\pi)^3 2E_1}\frac{d^3p_2}{(2\pi)^3 2E_2}\delta^4(m_X-p_1-p_2)(2\pi)^4 (-i)g_{2^+D^*D} p_X^\gamma \varepsilon_X^{\rho\eta} \epsilon_{\gamma\rho o\sigma}p_1^o p_{2\eta}i g_{J/\psi D^*D^*} \varepsilon^{*\kappa}_\psi \nonumber\\
&& [(q_{\kappa}-p_{1\kappa})(-g^{\sigma\tau}+\frac{p_1^\sigma p_{1}^\tau}{m_{D_S^*}^2})(-g_{\tau\beta}+\frac{q_\tau q_\beta}{m_{D_s^*}^2})+\nonumber\\
&&(-p_{\psi}^\tau-q^{\tau})(-g_{\beta\kappa}+\frac{q_\beta q_\kappa}{m_{D_s^*}^2})(-g_{\tau}^{\ \sigma}+\frac{p_1^\sigma p_{1\tau}}{m_{D_s^*}^2})+\nonumber\\
&&(p_{\psi}^\tau+p_{1}^\tau)(-g_{\tau\beta}+\frac{q_\tau q_\beta}{m_{D_s^*}^2})(-g_{\kappa}^{\ \sigma}+\frac{p_1^\sigma p_{1\kappa}}{m_{D_s^*}^2})]\nonumber\\
&&(-2 i)f_{D^*DV}\epsilon^{\mu\nu\alpha\beta}\varepsilon^*_{\phi\nu}(-i)p_{\phi\mu} (-ip_{2\alpha}-iq_\alpha)\frac{i}{q^2-m_{D_s^*}^2}{\cal F}^2(M_{D_s^{*+}},q)
\end{eqnarray}

\begin{eqnarray}
A_{2-m}=&&\frac{1}{2}\int \frac{d^3 p_1}{(2\pi)^3 2E_1}\frac{d^3p_2}{(2\pi)^3 2E_2}\delta^4(m_X-p_1-p_2)(2\pi)^4 ig_{2^+D^*D^*} \varepsilon_X^{\kappa\lambda}(-i)g_{J/\psi DD^*}\epsilon^{\mu\nu\alpha\beta}p_{\psi\mu}\varepsilon^*_{\psi\nu}p_{1\alpha}\nonumber\\
&&(-i)2 f_{D^*DV}\epsilon^{\rho\sigma\tau\eta}(q_\tau+p_{2\tau})\varepsilon^*_{\phi\sigma}p_{\phi\rho}(-g_{\kappa\beta}+\frac{p_{1\kappa}p_{1\beta}}{m_{D_S^*}^2}) (-g_{\lambda\eta}+\frac{p_{2\lambda}p_{2\eta}}{m_{D^*_s}})\nonumber\\
&&\frac{i}{q^2-m_{D_s}^2}{\cal F}^2(M_{D_s^{-}},q)
\end{eqnarray}

\begin{eqnarray}
A_{2-o}=&&\frac{1}{2}\int \frac{d^3 p_1}{(2\pi)^3 2E_1}\frac{d^3p_2}{(2\pi)^3 2E_2}\delta^4(m_X-p_1-p_2)(2\pi)^4 i g_{2^+D^*D^*} \varepsilon_X^{\rho\sigma} g_{J/\psi D^*D^*}\varepsilon_{\psi}^{*\nu}\nonumber\\
&&[(i p_{1\nu}-i q_{\nu})(-g_{\mu\rho}+\frac{p_{1\mu}p_{1\rho}}{m_{D_s}^*2})(-g_{\alpha\mu}+\frac{q_\alpha q_\mu}{m_{D_s}^*2})+\nonumber\\
&&(-i p_{\psi\mu}-i p_{1\mu})(-g_{\rho\nu}+\frac{p_{1\rho} p_{1\nu}}{m_{D^*_s}^2})(-g_{\alpha\mu}+\frac{q_\alpha q_\mu}{m_{D^*_s}^2})\nonumber\\
&&+(-g_{\mu\rho}+\frac{p_{1\mu} p_{1\rho}}{m_{D^*_s}^2})(-g_{\nu\alpha}+\frac{q_\alpha q_\nu}{m_{D^*_s}^2})(i q_\mu+i p_{\psi\mu})]\nonumber\\
&&[-g_{D^*D^*V}\varepsilon^{*\beta}_\phi (-i q_\beta-i p_{2\beta}(-g_\sigma^{\ \alpha}+\frac{p_{2\sigma}p_2^\alpha}{m_{D^*_s}^2}))\nonumber\\
&&-4 f_{D^*D^*V}(-i p_\phi^{\beta} \varepsilon_\phi^{\alpha*}+i p_\phi^{\alpha} \varepsilon_\phi^{\beta*})(-g^{\sigma\beta}+\frac{p_2^\sigma p_2^\beta}{m_{D_s^*}^2})]\frac{i}{q^2-m_{D_s^*}^2}{\cal F}^2(M_{D_s^{*-}},q)
\end{eqnarray}
$\varepsilon_X$ denote the
polarization tensors of $X(4350)$ which can be constructed from the
polarization vector of massive vector bosons as follows:
\begin{eqnarray}
\epsilon^\lambda_{\mu\nu}=\{\sqrt{2} \varepsilon^+_\mu \varepsilon^+_\nu, (\varepsilon^+_\mu \varepsilon^0_\nu + \varepsilon^0_\mu \varepsilon^+_\nu), \frac{1}{\sqrt{3}}(\varepsilon^+_\mu \varepsilon^-_\nu + \varepsilon^-_\mu \varepsilon^+_\nu -2 \varepsilon^0_\mu \varepsilon^0_\nu),(\varepsilon^-_\mu \varepsilon^0_\nu + \varepsilon^0_\mu \varepsilon^-_\nu), \sqrt{2} \varepsilon^-_\mu \varepsilon^-_\nu\}
\end{eqnarray}
The polarization tensor is traceless, transverse and orthogonal:$(\varepsilon^\lambda)^\mu_\mu=0$ ,$k^\mu \varepsilon^\lambda_{\mu\nu}=0$, $\varepsilon^{\lambda,\mu\nu}\varepsilon^{\lambda'*}_{\mu\nu}=2\delta^{ss'}$,
$\lambda$ could be 2 or 1 or 0 or -1 or -2. And
\begin{eqnarray}
\sum^5_{\lambda=1}\varepsilon^\lambda_{\mu\nu} \varepsilon^{\lambda *}_{\alpha\beta} = B_{\mu\nu,\alpha\beta}(k) .
\end{eqnarray}
\begin{eqnarray}
B_{\mu\nu,\alpha\beta}(k)=&&(g_{\mu\alpha}-\frac{k_\mu k_\alpha}{m^2})(g_{\nu\beta}-\frac{k_\nu k_\beta}{m^2})+(g_{\mu\beta}-\frac{k_\mu k_\beta}{m^2})(g_{\nu\alpha}-\frac{k_\nu k_\alpha}{m^2})\nonumber\\
&& -\frac{2}{3}(g_{\mu\nu}-\frac{k_\mu k_\nu}{m^2})(g_{\alpha\beta}-\frac{k_\alpha k_\beta}{m^2})
\end{eqnarray}
It is obvious that $k^\mu B_{\mu\nu\alpha\beta}=0$, $B^\mu_{\mu,\alpha\beta}=0$, for details, see \cite{9811350v4}(The Appendix part)\\
Similar with $X(3915)$, $A_{2-b}=A_{2-a}$, $A_{2-c}=A_{2-d}$, $A_{2-f}=A_{2-e}$, $A_{2-h}=A_{2-g}$, $A_{2-j}=A_{2-i}$, $A_{2-l}=A_{2-k}$, $A_{2-n}=A_{2-m}$, $A_{2-p}=A_{2-o}$, so:
\begin{eqnarray}
M(4350)=2(A_{2-a}+A_{2-c}+A_{2-e}+A_{2-g}+A_{2-j}+A_{2-l}+A_{2-m}+A_{2-o})
\end{eqnarray}
We are unable to calculate the width of $X(4350)\rightarrow J/\psi \phi$ since we can't decide the relative phase between $g_{2^+DD}$, $g_{2^+D^*D}$ and $g_{2^+D^*D^*}$. But we can check which channel is dominance by assuming $X(4350)$ decay to $J/\psi$ and $\phi$ in one of the three way. We show the result in Figure 4. It is clear that $D_s^{*+} D_s^{*-}$ is the dominance channel.

\begin{figure}[htb]
\centering \scalebox{0.8}{\includegraphics{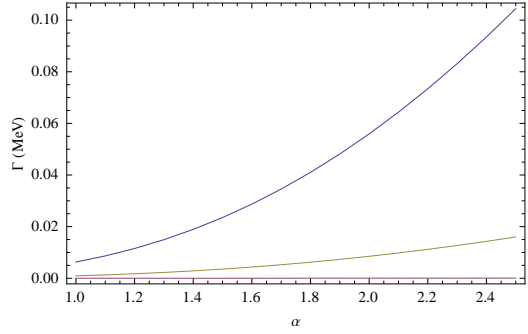}}
\caption{(Color on line) The decay rate for $X(4350)\rightarrow J/\psi+\phi$ via $D_s^+ D_s^-$ (green line), $D^*_s D_s$ (blue line), and $D^{*+}_s D^{*-}_s$ (red line).}
\end{figure}

\section{Summary}
Using 3p0 model to fix the coupling constants, we studied the hidden charm decays of $X(3915)\rightarrow J/\psi \omega$ and $X(4350)\rightarrow J/\psi \phi$ by assuming $X(3915)$ and $X(4350)$ are P-wave charmonium states. It seems that this assumption works well for $X(4350)\rightarrow J/\psi \phi$. While things are not very good for X(3915) since the experimental results implied that $\Gamma(X(3915)\rightarrow J/\psi  \omega)$ should be around 1 MeV \cite{Belle 1}. Thus $X(3915)$ may not be regarded as a pure charmonium state. This has been also pointed out \cite{0910.3138} and \cite{0912.5061}.

\begin{newpage}
\begin{center}
{\bf Appendix: The coupling constant}
\end{center}
\setcounter{equation}{00}
\renewcommand{\theequation}{A.\arabic{equation}}

The 3p0 model use to do a good job in calculating the OZI allowed decay width in meson's strong decay. So we can use it to study the amplitude of $X(3915) \rightarrow D^+ D^-$ and $X(4350) \rightarrow D_s$ and $D_s^*$, then decide the coupling constants $g_{0^+DD}$, $g_{2^+DD}$, $g_{2^+D^*D}$ and $g_{2^+D^*D^*}$.
\begin{figure}[htb]
\centering \scalebox{0.5}{\includegraphics{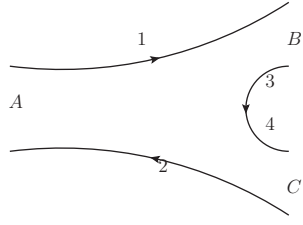}}
\caption{The diagram for meson A decay into meson B and C}
\end{figure}
When a heavy quark meson like $X(4350)$ or $X(3915)$ decays, a pair of light quark-antiquark created from the vacuum with the vacuum's quantum number $0^{++}$. The pair should be singlet in color, flavor, of zero momentum and zero total angular momentum. The parity should be positive. So $L=1$, $S=1$. Then the quark-antiquark separate and enter into different mesons. See Figure 5. The transition matrix T is defined as:
\begin{eqnarray}
 S =1-2\pi i \delta(E_f-E_i) T
\end{eqnarray}
 and express T in the form of nonrelativistic limit:
 \begin{eqnarray}
 T=-\int d^3p_3 d^3p_4[3\gamma \delta(p_3+p_4)\sum_m <1,m;1,-m|0,0> |k| Y_{1m}(k)(\chi_{1,-m}\,\varphi_0\,\omega_0)]b^\dag(p_3) d^\dag(p_4),
 \end{eqnarray}
 $\varphi_0$ is for $SU(3)_F$ singlet ($\varphi_0=-(u \bar u+d \bar d+ s \bar s)/\sqrt{3}$), $\omega_0$ for color-singlet and $\chi_{1,-m}$ for triplet state of spin \cite{book}. $Y_{1,m}$ is the spheric harmonic function, reflecting the $L=1$ orbital angular momentum of the pair. The quark-antiquark pair is created by $b^\dag(p_3)$ and $d^\dag(p_4)$. $k=(m_i k_j-m_j k_i)/(m_i+m_j)$ is the relative momentum between the quark and the antiquark within a meson. $\gamma$ is a dimensionless constant that correspond to the strength of the transition. The state of the meson A can be written as:
 \begin{eqnarray}
 |A(J,M_{J},L,S)>\;=&&\sqrt{2E_A}\sum_{M_L,M_S}<L\,M_L\,S\,M_S|J\,M_J>\int d^3p_1 d^3p_2 \delta(k_A-p_1-p_2)\nonumber\\
 &&\Psi_{nLM_L}(k) \chi_{SM_S} \varphi_A \omega_A b^\dag_{p1} d^\dag_{p2}|0>
 \end{eqnarray} The states of B and C are similar.
  $\Psi_{nLM_L}(k)$ is the spatial part of a meson's function. In this work, we use the simple harmonic oscillator (SHO) potential as the interaction potential function inside a meson. So $\Psi_{nLM_L}(k)$ could be expressed as \cite{1008.0246}:
\begin{eqnarray}
\Psi_{nLM_L}(k)=(-1)^n(-i)^L R^{3/2} \sqrt{\frac{2 n!}{\Gamma(n+L+3/2)}}(kR)^L exp(-2k^2R^2)L_n^{L+1/2}(k^2 R^2)Y_{L M_L}
\end{eqnarray}

 $L_n^{L+1/2}(k^2R^2)$ is the Laguerre polynomial. R is a parameter comes from the potential function. If we take the center of mass frame of meson A, the element of the T matrix can be expressed as:
 \begin{eqnarray}
 <BC|T|A>=&&\sqrt{8E_AE_BE_C}\gamma \sum_{\mbox{\tiny$\begin{array}{c}
M_{L_A},M_{S_A},\\
M_{L_B},M_{S_B},\\
M_{L_C},M_{S_C},m\end{array}$}}
<L_AM_{L_A}S_AM_{S_A}|J_AM_{J_A}> \nonumber\\
 && <L_BM_{L_B}S_BM_{S_B}|J_BM_{J_B}><L_CM_{L_C}S_CM_{S_C}|J_CM_{J_C}> <1\;m;1\;-m|0\;0>\nonumber\\
 &&<\varphi_B\varphi_C|\varphi_A\varphi_0><\chi_{S_BM_{S_B}}\chi_{S_CM_{S_C}}| \chi_{S_AM_{S_A}}\chi_{1\,-m}>\nonumber\\
 &&<\omega_B(1,3)\omega_C(2,4)|\omega_A(1,2) P(3,4)>I^{M_{L_A},m}_{M_{L_B},M_{L_C}}(k)
 \end{eqnarray}

 The spatial integral is:
 \begin{eqnarray}
 I^{M_{L_A},m}_{M_{L_B},M_{L_C}}(k)=&&\int d^3p_1d^3p_2d^3p_3d^3p_4\delta(p_1+p_2)\delta(p_3+p_4)\delta(k_B-p_1-p_3)\delta(k_C-p_2-p_4)\nonumber\\
 &&\times\Psi^*_{n_BL_BM_{L_B}}(p_1,p_3)\Psi^*_{n_CL_CM_{L_C}}(p_2,p_4)\Psi_{n_AL_AM_{L_A}}(p_1,p_2)|\frac{p_3-p_4}{2}|Y_{1m}
 \end{eqnarray}

 In this way, the amplitudes of $X(3915)\rightarrow D^{+}D^-$ or $D^0\bar{D^0}$ and $X(4350)\rightarrow D_s D_s^*$ can be expressed like:
 \begin{eqnarray}
 {\cal M}(X(3915)\rightarrow D^{+}D^-)=\frac{\sqrt 2}{3\sqrt3} \sqrt{E_AE_BE_C} \gamma [2I^{1,-1}_{0,0}-I_{0,0}^{0,0}]\\
 {\cal M}(X(3915)\rightarrow D^0\bar{D^0})=\frac{\sqrt 2}{3\sqrt3} \sqrt{E_AE_BE_C} \gamma [2I^{1,-1}_{0,0}-I_{0,0}^{0,0}]\\
 {\cal M}(X(4350)\rightarrow D_sD_s)=\frac{2}{3\sqrt{15}} \sqrt{E_AE_BE_C} \gamma [I^{1,-1}_{0,0}+I_{0,0}^{0,0}]\\
 {\cal M}(X(4350)\rightarrow D_sD_s^*)=\frac{2}{3\sqrt{10}} \sqrt{E_AE_BE_C} \gamma [I^{1,-1}_{0,0}+I_{0,0}^{0,0}]\\
 {\cal M}(X(4350)\rightarrow D_s^*D_s^*)=\frac{2\sqrt2}{9} \sqrt{E_AE_BE_C} \gamma [2I^{1,-1}_{0,0}-I_{0,0}^{0,0}]
 \end{eqnarray}

 $\gamma=6.3$ \cite{prd 43.1679} for the quark-antiquark pair creation of $u\bar{u}$ and $d\bar{d}$, $\gamma=6.3/\sqrt{3}$ for $s\bar{s}$ \cite{plb 72.57}. We take $m_c=1.6 \,GeV$, $m_u=m_d=0.22\,GeV$, $m_s=0.419 \,GeV$. According to \cite{prd 43.1679}, $R_D=1.52 \,GeV^{-1}$, $R_{D_s}=1.41\, GeV^{-1}$, $R_{D^*_s}=1.69\, GeV^{-1}$. It was decided in \cite{Xiang Liu 1} that $R_{X(3915)}=1.80\sim1.99 \,GeV^{-1}$, $\,1.92 GeV^{-1}$ for the central value of decay width and $R_{X(4350)}=1.8\sim3.0 \,GeV^{-1}$, central value is not available. We will choose $R_{X(3915)}=1.92 \,GeV^{-1}$, $R_{X(4350)}=1.90\, GeV^{-1}$ in this work. The decay rates could be calculated in this way: $\Gamma(X(3915)\rightarrow D^+D^-)= 8.446$MeV, $\Gamma(X(4350)\rightarrow D^+_sD^-_s)= 0.518$MeV, $\Gamma(X(4350)\rightarrow D^{*+}_sD^-_s)= 2.915$MeV, $\Gamma(X(4350)\rightarrow D^{*+}_sD^{*-}_s)= 1.282$MeV. Then we find the numerical result  for the coupling constant is \begin{eqnarray}
\left\{ \begin{array}{ll}
|g_{0^+DD}|=2370 MeV   \\
|g_{2^+DD}|=0.002 MeV^{-1}   \\
|g_{2^+D^*D}|=3.30\times 10^{-7} MeV^{-2}  \\
|g_{2^+D^*D^*}|=700 MeV
\end{array} \right .
\end{eqnarray}
\end{newpage}

\section*{Acknowledgments}
I would like to thank Dao-Neng Gao, Gui-Jun Ding and Jia-Feng Liu for helpful discussions. This work is support in part by the NSF of China under grant No. 10775124 and 11075149.

\end{document}